\documentclass[12pt,a4j]{article}
\usepackage[dvips]{color,graphicx}
\usepackage{amsmath,wrapfig,amsfonts}
\usepackage{natbib}
\newcommand{\bm}[1]{\mbox{\boldmath{$#1 $}}} 
\begin{document}
\baselineskip=20pt
\begin{center}
{\Large\textbf{Model selection criteria for \\ nonlinear mixed effects modeling}}\\~\\
\large{Hidetoshi Matsui\footnote{Faculty of Mathematics, Kyushu University. 744 Motooka, Nishi-ku, Fukuoka 819-0395, Japan.  \\E-mail: hmatsui@math.kyushu-u.ac.jp}\\
{\it Kyushu University}
}
\\
\end{center}
\begin{abstract}
We consider constructing model selection criteria for evaluating nonlinear mixed effects models via basis expansions.  
Mean functions and random functions in the mixed effects model are expressed by basis expansions, then they are estimated by the maximum likelihood method.  
In order to select numbers of basis we derive a Bayesian model selection criterion for evaluating nonlinear mixed effects models estimated by the maximum likelihood method.  
Simulation results shows the effectiveness of the mixed effects modeling.  \\

\end{abstract}
\section{introduction}
Mixed effects modeling is an effective technique for analyzing data with a complex structure, and is an extension of traditional linear models that allow for the incorporation of random effects.  
\cite{LaWa1982} applied the mixed effects model to the analysis of repeated measures data, and developed the methodology for formulation and fitting of it.   
It can be easily applied even if the data have few observational points or irregularly spaced points, by analyzing the complete set of data at one time.  
Analysis of longitudinal data via the mixed effects modeling has been widely studied, especially in medical science \citep{ArBeMa2008, FiLaWa2012}.  

\cite{BrRi1998} extended the linear mixed effects model to that with nonlinear structure by approximating individual curves as spline functions, enabling us to construct more flexible models.  
\cite{RiWu2001} assumed a more general structure for the covariance function and then computed eigenfunctions which provide insights into individual curves.  
The estimated curves can also be considered as a set of functional data \citep{RaSi2005}.  
The basic idea behind the functional data analysis is to represent observed longitudinal data as smooth functions and then treat each of them as individual data.  
Therefore we can apply further analyses to the estimated curves such as functional version of principal component analysis or regression analysis.  

The mixed effects model is estimated by the framework of the maximum likelihood method.  
 \cite{LaWa1982} estimated the linear mixed effects model by the EM algorithm under the assumption that the variance structure is unknown.  
When constructing the mixed effects model it is a crucial issue to select variables since it directly leads to the prediction accuracy.  
One of the solution for the issue is the selection via model selection criteria.  
For the linear mixed effects model, \cite{VaBl2005} derived a conditional AIC which apply the effective degrees of freedom for the linear mixed effects model, and afterward \cite{LiWuZo2008} extended it so that it can be used in more general conditions.  
Further description about the model selection criteria for the linear mixed effects model are given in \cite{BuAn2002}.    

Similary, in the nonlinear setting we also need to select select an optimal model capturing both mean functions and random functions.  
Moreover, we should select the model more carefully since selection of a few tuning parameters can control the model constructed from the complete data set.  

We introduce some model selection criteria, derived from information theory and a Bayesian approach, for evaluating the nonlinear mixed effects model estimated by the maximum likelihood method.  
Especially we derive an improved version of Bayesian model selection criterion \citep{Sc1978} by applying the result of \cite{KoAnIm2004}.  
These criteria can be used even if the covariance structure of the random effects and the error variance are unknown.  
In order to investigate the effectiveness of the proposed criteria simulation studies are conducted.  

This paper is organized as follows.  
Section 2 introduces nonlinear mixed effects models based on basis expansions.  
In Section 3 we describe the maximum likelihood procedure for estimating the nonlinear mixed effects model, assuming that parameters and the structure of the variance are unknown.  
Section 4 shows some model selection criteria for evaluating the nonlinear mixed effects model.  
Simulation examples and real data analysis are investigated in Section 5 and finally concluding remarks are given in Section 6.  
\section{Nonlinear mixed effects model via basis expansions}
Suppose we have repeated measurement data $\{(t_{\alpha i}, x_{\alpha i});$ $\alpha=1,\ldots,$ $n,$ $i=1,\ldots,$ $N_\alpha\}$, where $t_{\alpha i}$ is the $i$-th time point for the $\alpha$-th subject and $x_{\alpha i}$ is the observed value at $t_{\alpha i}$.  Then we consider the following model
\begin{align*}
x_{\alpha i} = m(t_{\alpha i}) + r_\alpha(t_{\alpha i})+ \varepsilon_{\alpha i},
\end{align*}
where $m(t)$ is an overall mean function, $r_\alpha(t)$ are random functions and $\varepsilon_{\alpha i}$ are noise variables.  In many works on mixed effects modeling, $m(t_{\alpha i})$ and $r_\alpha(t_{\alpha i})$ are expressed by linear combination of $\bm t_{\alpha} = (t_{\alpha 1},$ $\ldots,$ $t_{\alpha N_\alpha})^T$ and known vectors respectively.  On the other hand, we assume that $m(t)$ and $r_\alpha(t)$ can be expressed as linear combinations of $m_f$ basis functions $\phi_k^f(t)$ and $m_r$ basis functions $\phi_l^r(t)$ respectively \cite{RiWu2001}, that is, $x_{\alpha i}$ are represented as
\begin{align}
x_{\alpha i} = \sum_{k=1}^{m_f}\beta_k \phi_k^{f}(t_{\alpha i})+\sum_{l=1}^{m_r}\gamma_{\alpha l} \phi_l^{r}(t_{\alpha i})+\varepsilon_{\alpha i},\label{mm1}
\end{align}
where $\beta_k$ and $\gamma_{\alpha l}$ are coefficients.  
The equation (\ref{mm1}) can be expressed using vector and matrix notation as follows:
\begin{align*}
\bm x_\alpha = \Phi_\alpha^f\bm\beta+\Phi_\alpha^r\bm \gamma_\alpha+\bm\varepsilon_\alpha,
\end{align*}
where $\bm x_\alpha=(x_{\alpha 1},\ldots,$ $x_{\alpha N_\alpha})^{T}$, $\Phi_\alpha^f=(\phi_k^f(t_{\alpha i}))_{ik}$, $\Phi_\alpha^r=(\phi_l^r(t_{\alpha i}))_{il}$, $\bm\beta$ $=(\beta_1,\ldots,$ $\beta_{m_f})^{T}$, $\gamma_\alpha=(\gamma_{\alpha i},\ldots,$ $\gamma_{\alpha m_r})^{T}$ and $\bm\varepsilon_\alpha=(\varepsilon_{\alpha 1},\ldots,$ $\varepsilon_{\alpha N_\alpha})^{T}$ and we make the following assumptions:
\begin{align*}
\bm\gamma_\alpha\sim N_{m_r}(\bm 0,\Gamma),~~~~~~\bm\varepsilon_\alpha\sim N_{N_\alpha}(\bm 0,\sigma_\varepsilon^2 I).
\end{align*}

A typical choice of basis functions is Fourier series or $B$-splines \citep{de2001, ImKo2003}, and Gaussian radial basis functions are also used \citep{Bi1995, AnKoIm2008}.  
Here we assume that $\phi_k^f(t)$ and $\phi_l^r(t)$ are $B$-splines of degree 3.  
Suppose we have equally spaced knots $\tau_k$ such that $\tau_1<\cdots<\tau_{r+1}=\min(t_{\alpha i})<\cdots<\tau_{p+1}=\max(t_{\alpha i})<\cdots<\tau_{p+r+1}$, then $B$-spline functions of degree 0 are defined by
\begin{align*}
B_j(t;0)=\left\{\begin{array}{ll}
1 & (\tau_j\le t<\tau_{j+1}),\\
0 & (\mathrm{otherwise}).
\end{array}\right.
\end{align*}
Then $B$-spline functions of degree $r$ are formed using the following sequential equations:
\begin{align*}
B_j(t;r)=\frac{t-\tau_j}{\tau_{j+r}-\tau_j}B_j(t;r-1)+\frac{\tau_{j+r+1}-t}{\tau_{j+r+1}-\tau_{j+1}}B_{j+1}(t;r-1).
\end{align*}
We apply the functions of degree 3 $B_k(t; 3)$ and $B_l(t; 3)$ to basis functions $\{\phi_k^f(t);$ $k=1,\ldots, m_f\}$ and $\{\phi_l(t);$ $l=1,\ldots, m_r\}$ respectively.  

The nonlinear mixed effects model can then be expressed as
\begin{align}
f(\bm x_\alpha|\bm t_\alpha; \bm\theta) =&
\frac{1}{(2\pi)^{N_\alpha/2}|\sigma^2_\varepsilon I_{N_\alpha}+\Phi^r_\alpha\Gamma \Phi^{r^{T}}_\alpha|^{1/2}}\nonumber\\
& ~~\times \exp\left\{
-\frac{1}{2}(\bm x_\alpha-\Phi^f_\alpha\bm\beta)^{T}(\sigma^2_\varepsilon I_{N_\alpha}+\Phi^r_\alpha \Gamma \Phi^{r^{T}}_\alpha)^{-1}(\bm x_\alpha-\Phi^f_\alpha\bm\beta)\right\} \label{mm-pdf},
\end{align}
where $\bm\theta=\{\bm\beta^T, ({\rm vech}\Gamma)^T, \sigma^2_\varepsilon\}^T$ is a parameter vector and ${\rm vech}\Gamma$ denotes an operator that transforms $m_r(m_r+1)/2$ upper triangular elements of $\Gamma$ into a vector.  
\section{Estimation}
We consider estimating the nonlinear mixed effects model (\ref{mm-pdf}) by the maximum likelihood method.  
The log-likelihood function of the model is given by
\begin{align}
l(\bm\theta) =& -\frac{1}{2}\sum_{\alpha=1}^n \left\{N_\alpha\log(2\pi) + \log|\sigma^2_\varepsilon I_{N_\alpha} + \Phi^r_\alpha \Gamma \Phi^{r^{T}}_\alpha|\right\}\nonumber\\
& ~~-\frac{1}{2}\sum_{\alpha=1}^n \left\{(\bm x_\alpha -\Phi^f_\alpha\bm\beta)^{T}(\sigma^2_\varepsilon I_{N_\alpha} + \Phi^r_\alpha \Gamma \Phi^{r^{T}}_\alpha)^{-1}(\bm x_\alpha -\Phi^f_\alpha \bm\beta)\right\}.\label{mm-loglike}
\end{align}
Following subsections describe how to obtain the estimator of $\bm\theta$ for cases where variance parameters $\Gamma$ and $\sigma_\varepsilon^2$ are known and unknown.  
\subsection{Known variances}
When both $\Gamma$ and $\sigma_\varepsilon^2$ are known, the maximum likelihood estimator of $\bm\beta$ is easily obtained from (\ref{mm-loglike}):
\begin{align*}
\hat{\bm\beta}=\left(\sum_{\alpha=1}^n\Phi_\alpha^{f^{T}}W^{-1}\Phi_\alpha^f\right)^{-1}\sum_{\alpha=1}^n\Phi_\alpha^{f^{T}}W^{-1}\bm x_\alpha, 
\end{align*}
where $W=\sigma^2_\varepsilon I_{N_\alpha}+\Phi_\alpha^r\Gamma\Phi_\alpha^{r^{T}}$.  \cite{LaWa1982} derived predictors $\hat{\bm\gamma}_\alpha$ by using an extension of the Gauss-Markov theorem \citep{Ha1976}, given by
\begin{align}
\hat{\bm\gamma}_\alpha=\Gamma\Phi_\alpha^{r^{T}}W^{-1}(\bm x_\alpha-\Phi_\alpha^f\hat{\bm\beta}).
\label{mm-u}
\end{align}
Their estimates or predictors are given as BLUP (Best Linear Unbiased Prediction) estimators which has minimum variance in unbiased estimates or predictors of $\bm\beta$ or $\bm\gamma_\alpha$.  
Theory of BLUP is discussed by \cite{Ro1991}.  
\subsection{Unknown variances}  
It is unnatural that the variance parameters $\Gamma$ and $\sigma_\varepsilon^2$ are known, and thus we assume that these are unknown.  
However, when both of them are unknown it is difficult to derive maximum likelihood estimators $\hat\sigma^2_\varepsilon$, $\hat\Gamma$ and $\hat{\bm\beta}$ analytically from the log-likelihood (\ref{mm-loglike}).  
Alternatively, they can be estimated via the EM algorithm, considering $\bm \gamma_\alpha$ as latent variables \citep{LaWa1982}.  
If $\bm\gamma_\alpha$ were observed, the density function of $\bm x_\alpha$ would be given by
\begin{align*}
f(\bm x_\alpha | \bm t_\alpha, \bm\gamma_\alpha; \bm\theta) &=
\frac{1}{(2\pi)^{(N_\alpha+m_r)/2}\sigma^{N_\alpha}_\varepsilon|\Gamma|^{1/2}}\nonumber\\
&~\times\exp\left\{
-\frac{1}{2\sigma_\varepsilon^2}(\bm x_\alpha-\Phi^f_\alpha{\bm\beta}-\Phi^r_\alpha\bm \gamma_\alpha)^{T}(\bm x_\alpha-\Phi^f_\alpha{\bm\beta}-\Phi^r_\alpha\bm \gamma_\alpha)-\frac{1}{2}\bm \gamma_\alpha^{T}\Gamma^{-1}\bm \gamma_\alpha\right\},
\end{align*}
which can be regarded as a complete log-likelihood function.  
Considering a conditional expectation $Q(\bm\theta|\tilde{\bm\theta})$ $:=$ $\sum_\alpha E_{\gamma_\alpha}[\log f(\bm x_\alpha,\bm\gamma_\alpha;$ $\bm\theta)|\bm x_\alpha;$ $\tilde{\bm\theta}]$ with the current estimate $\tilde{\bm\theta}$, the parameter $\bm\theta$ is updated by maximizing $Q(\bm\theta|\tilde{\bm\theta})$ since the maximizer of (\ref{mm-loglike}) coincides with that of $Q(\bm\theta|\tilde{\bm\theta})$.  
The details of the EM algorithm is given in Appendix A.  
Replacing the unknown parameter $\bm\theta$ in (\ref{mm-loglike}) by its estimator $\hat{\bm\theta}=\{\hat{\bm\beta}^T, ({\rm vech}\hat\Gamma)^T, \hat\sigma^2_\varepsilon\}^T$, we obtain the nonlinear mixed effects model
\begin{align}
f(\bm x_\alpha|\bm t_\alpha; \hat{\bm\theta}) =&
\frac{1}{(2\pi)^{N_\alpha/2}|\hat\sigma^2_\varepsilon I_{N_\alpha}+\Phi^r_\alpha\hat\Gamma \Phi^{r^{T}}_\alpha|^{1/2}}\nonumber\\
&\times \exp\left\{
-\frac{1}{2}(\bm x_\alpha-\Phi^f_\alpha\hat{\bm\beta})^{T}(\hat\sigma^2_\varepsilon I_{N_\alpha}+\Phi^r_\alpha \hat\Gamma \Phi^{r^{T}}_\alpha)^{-1}(\bm x_\alpha-\Phi^f_\alpha\hat{\bm\beta})\right\}.\label{mm-estimate}
\end{align}
Moreover, predictors of $\bm \gamma_\alpha$ and $\bm x_\alpha$ are, respectively, given by
\begin{align}
\hat{\bm\gamma}_\alpha &= (\hat\sigma^2_\varepsilon \hat\Gamma^{-1} + \Phi_\alpha^{r^{T}}\Phi_\alpha^r)^{-1}\Phi^{r^{T}}_\alpha(\bm x_\alpha -\Phi^f_\alpha\hat{\bm\beta}), \label{mm-u2}\\
\hat{\bm x}_\alpha &= \Phi^f_\alpha\hat{\bm{\beta}}+\Phi^r_\alpha\hat{\bm\gamma}_\alpha.\nonumber
\end{align}
We can find that the predictor (\ref{mm-u2}) coincides with (\ref{mm-u}) by using matrix algebra.
\section{Model selection criteria}
The nonlinear mixed effects model $f(\bm x_\alpha|\bm t_\alpha;\hat{\bm\theta})$ estimated by the maximum likelihood method depends on numbers of basis functions $m_f$ and $m_r$.  
It is a crucial issue to determine them appropriately since only these parameters control the degrees of complexity of the model.  
We introduce some model selection criteria for evaluating nonlinear mixed effects models estimated by the maximum likelihood method, when variances are unknown.  

Akaike's information criterion \citep{Ak1974} for evaluating the model (\ref{mm-estimate}) is given by
\begin{align*}
{\rm AIC}=-2\sum_{\alpha=1}^n \log f(\bm x_\alpha|\bm t_\alpha;\hat{\bm\theta})+2p,
\end{align*}
where $p$ is the number of unknown parameters and is given by $p= m_f + m_r(m_r + 1)/2 + 1$.  

The Bayesian model selection criterion BIC \citep{Sc1978} for evaluating the model (\ref{mm-estimate}) is given by
\begin{align*}
{\rm BIC}=-2\sum_{\alpha=1}^n \log f(\bm x_\alpha|\bm t_\alpha;\hat{\bm\theta})+p\log n.
\end{align*}
\cite{KoAnIm2004} derived an improved version of Schwarz's BIC for regression models estimated by the maximum likelihood method.  Using this result, we derive an improved version of BIC based on the nonlinear mixed effects model via basis functions, which is given by
\begin{align}
{\rm BIC_I} = -2\sum_{\alpha=1}^n \log f(\bm x_\alpha|\bm t_\alpha;\hat{\bm\theta}) + p\{\log n - \log(2\pi)\} + \log |I(\hat{\bm\theta})|,
\label{BIC2}
\end{align}
where $I(\bm\theta)$ is a $p\times p$ matrix and whose elements are described in Appendix B.  We select $m_f$ and $m_r$ which minimize values of these criteria, and then consider the corresponding model to be the optimal model.  
\section{Numerical example}
Monte Carlo simulations are conducted to examine the effectiveness of nonlinear mixed effects modeling.  For simplicity, in this simulation the observational points are supposed to be the same for each individual.  

\begin{figure}[t] 
\begin{tabular}{cc}
\begin{minipage}{0.5\hsize}
\begin{center}
\includegraphics[width=7cm,height=7cm]{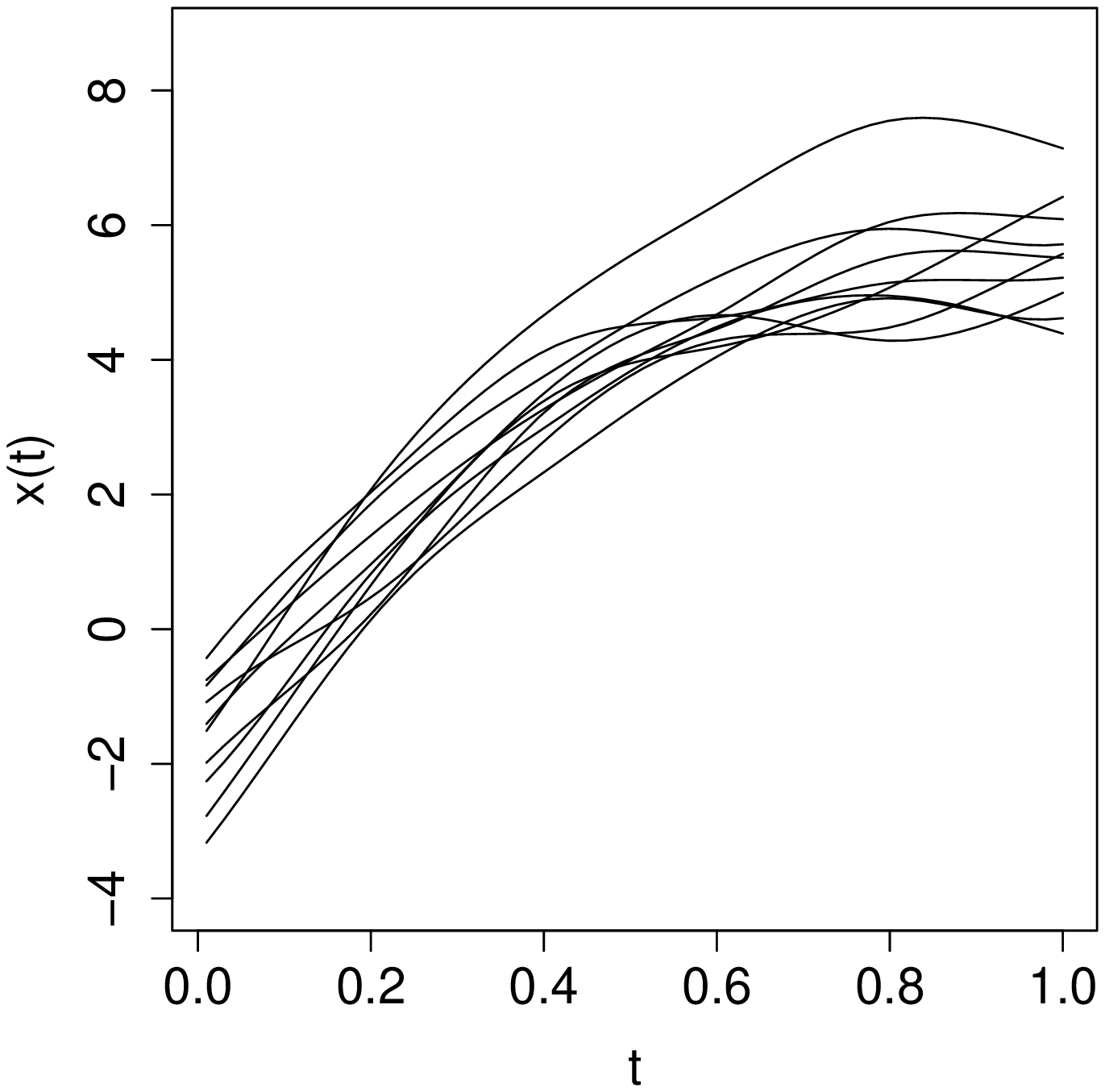}
\end{center}
\end{minipage}
\begin{minipage}{0.5\hsize}
\begin{center}
\includegraphics[width=7cm,height=7cm]{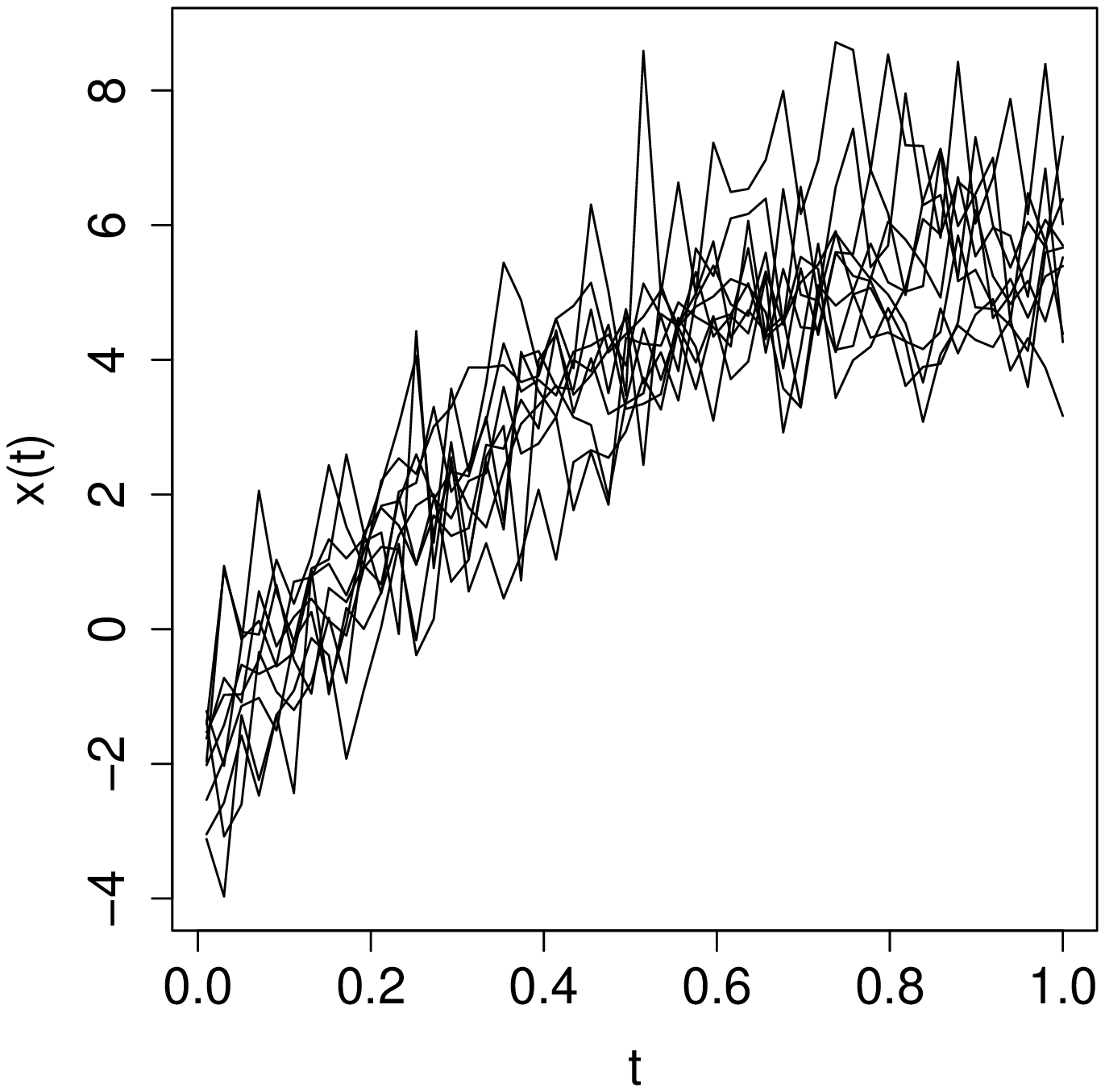}
\end{center}
\end{minipage}
\\
\begin{minipage}{0.5\hsize}
\begin{center}
\includegraphics[width=7cm,height=7cm]{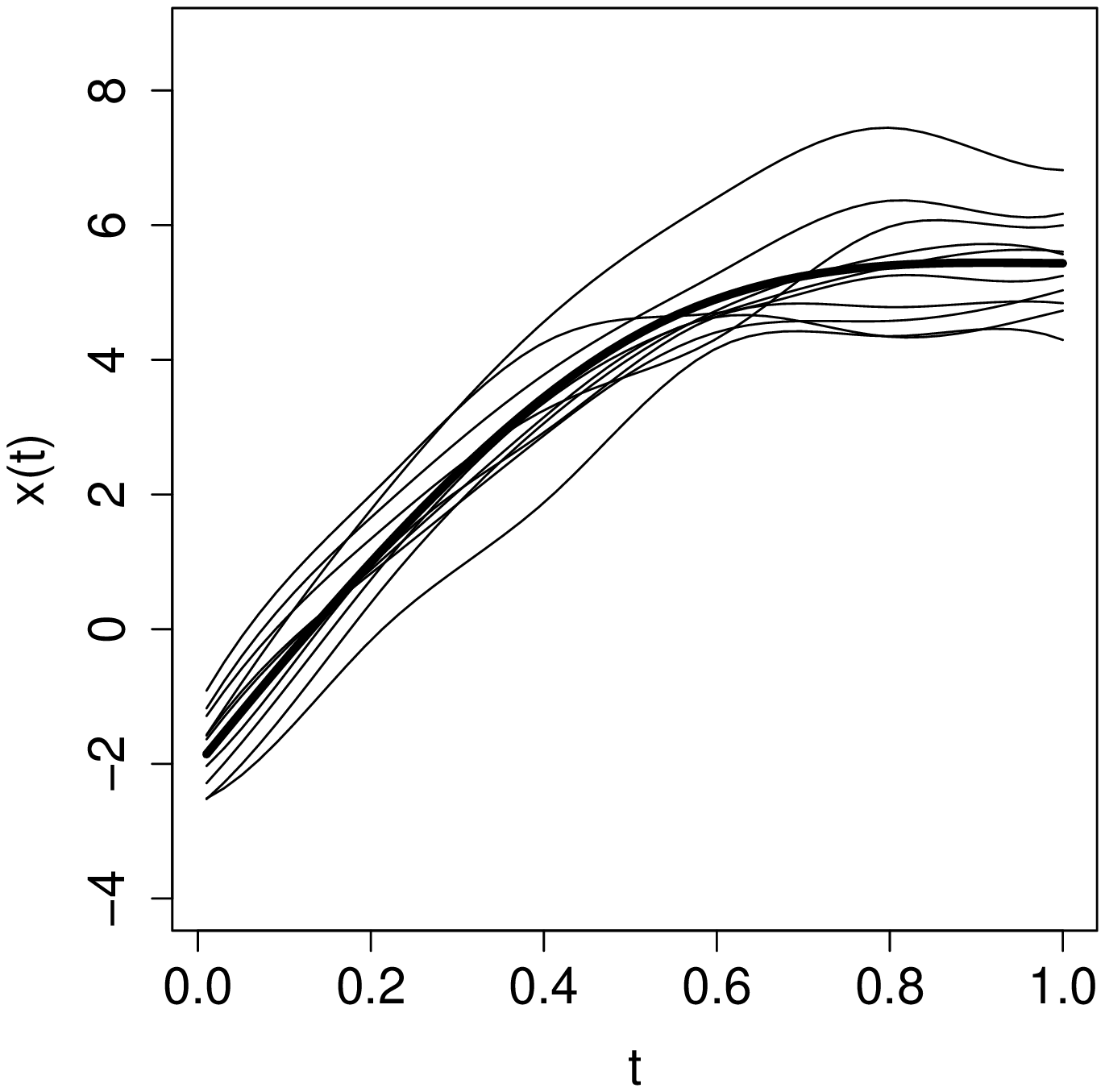}
\end{center}
\end{minipage}
\begin{minipage}{0.5\hsize}
\begin{center}
\includegraphics[width=7cm,height=7cm]{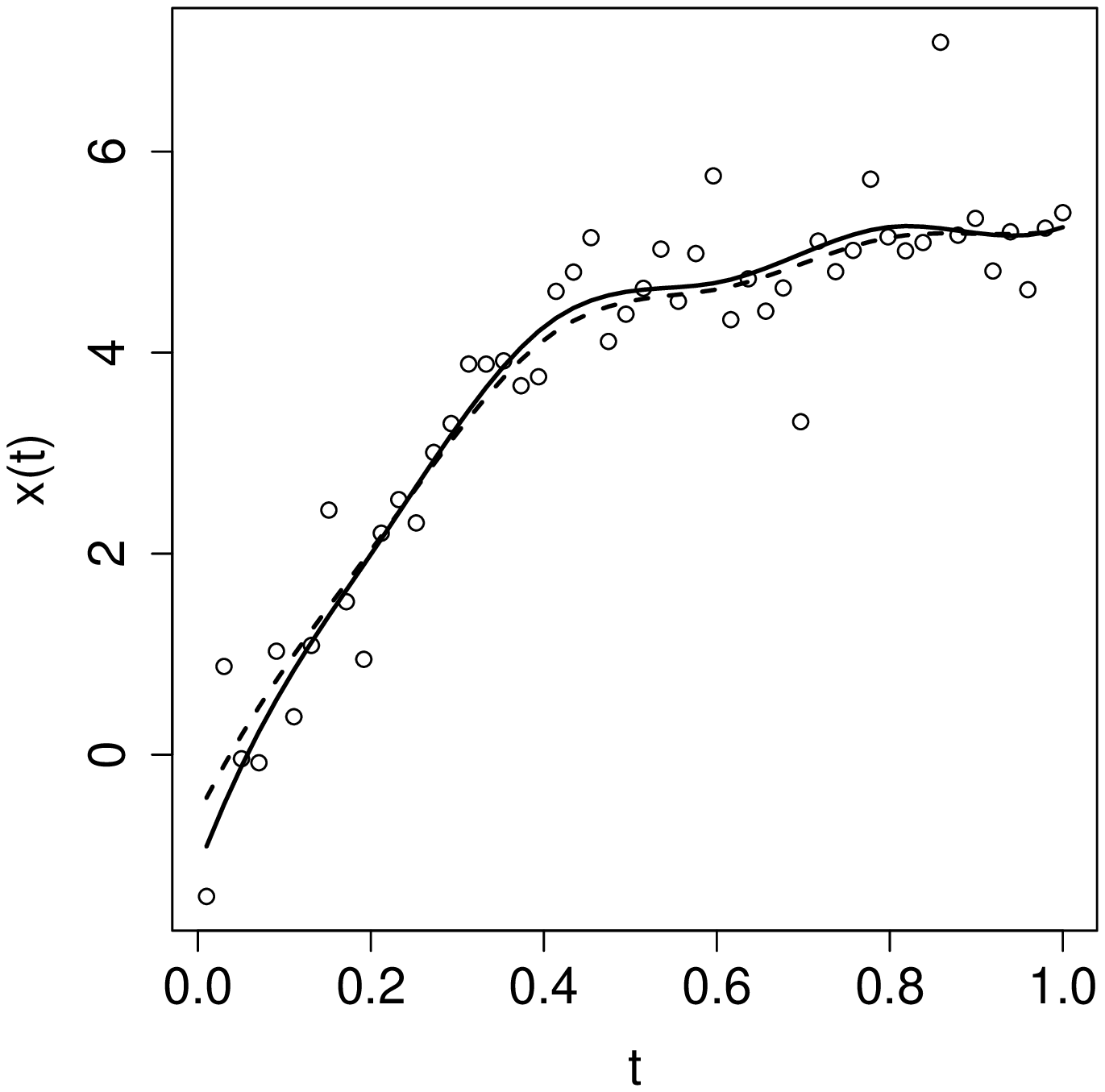}
\end{center}
\end{minipage}
\end{tabular}
\caption{10 examples of the simulation setting.  
Top left: True curves.  
Top right: Generated observations.  
Bottom left: Estimated curves with a fixed effect (thick line). 
Bottom right: An example of estimated curves.  
Points represent observations; the solid and dashed lines depict the true and estimated curves respectively.}
\label{fig:num-mixedmodel}
\end{figure}

First, we generated the $\alpha$-th observations $x_{\alpha i}$ at observational points $t_{\alpha i}$ $(\alpha=1,\ldots,$ $n,$ $i=1,\ldots,$ $50)$ using the following rule:  
\begin{align*}
&x_{\alpha i} = u_\alpha(t_{\alpha i}) + \varepsilon_{\alpha i},~~ \varepsilon_{\alpha i}\sim N(0, 0.1R_{x\alpha}^2),~~R_{x\alpha}=\max_i(u_\alpha(t_{\alpha i}))-\min_i(u_\alpha(t_{\alpha i})),\\
&u_\alpha(t_{\alpha i}) = \bm\beta^{T}\bm\phi^f(t_{\alpha i})+\bm\gamma_\alpha^{T}\bm\phi^r(t_{\alpha i}),~~~t_{\alpha i}=0.01+\frac{1-0.01}{50-1}(j-1),
\end{align*}
where $\bm\phi^f(t_{\alpha i})$ and $\bm\phi^r(t_{\alpha i})$ are $m_f$ and $m_r$ dimensional vectors of $B$-spline basis functions respectively.  Here we assume that $m_f=5$, $m_r=8$ and $\bm\beta=(-8,-2,6,5,7)^{T}$, and $\bm\gamma_\alpha$ are generated from $N_8(\bm 0, \Sigma_r)$ with $\Sigma_r=(0.5^{|j-k|})_{j,k}$.  We applied the nonlinear mixed effects modeling to the generated data, and then selected numbers of basis functions $m_f$ and $m_r$ using model selection criteria AIC, BIC and ${\rm BIC_I}$, thereby obtaining the estimator $\hat x_{\alpha i}$ for $n=30,$ $50,$ $100$.  We examined the simulations for 100 repetitions, then obtained the average mean squared error
\begin{align*}
{\rm AMSE} = \frac{1}{50n}\sum_{\alpha=1}^n\sum_{i=1}^{50}\left(\hat x_{\alpha i}-u_\alpha(t_{\alpha i})\right)^2.
\end{align*}
\begin{table}[t]
\caption{Averaged mean squared errors $(\times 10^2)$.}
\label{tab:mm-sim1}
\begin{center}
\begin{tabular}{cccc}
\hline
  \hline
$n$ & AIC & BIC & ${\rm BIC_I}$\\
\hline
 30 & 5.87 & 5.96 & 5.83\\
50 & 5.53 & 5.73 & 5.47\\
100 & 5.23 & 5.44 & 5.20\\
\hline
\end{tabular}
\end{center}
\caption{Frequency of the selected number of basis functions.  Bold numbers indicate correctly selected numbers of basis functions.  }
\label{tab:mm-sim2}
\begin{center}
\begin{tabular}{cccccccccc}
\hline
  \hline
 & $n$ & Criterion  & \multicolumn{7}{c}{Selected number of basis}\\
&& & 4 & 5 & 6 & 7 & 8 & 9 & 10 \\
\hline
$m_f$ & 30 & AIC & 4 & \bf 69 & 17 & 2 & 4 & 3 & 1\\
&&BIC & 10 & \bf 81 & 8 & 1 & 0 & 0 & 0\\
&&${\rm BIC_I}$ & 36 & \bf 60 & 4 & 0 & 0 & 0 & 0\\
&50 & AIC & 0 & \bf 62 & 18 & 7 & 4 & 5 & 4\\
&&BIC & 2 & \bf 82 & 13 & 3 & 0 & 0 & 0\\
&&${\rm BIC_I}$ & 3 & \bf 77 & 11 & 2 & 2 & 4 & 1\\
&100 & AIC & 0 & \bf 69 & 15 & 6 & 3 & 3 & 4\\
&&BIC & 0 & \bf 91 & 8 & 1 & 0 & 0 & 0\\
&&${\rm BIC_I}$ & 0 & \bf 72 & 15 & 4 & 3 & 2 & 4\\
\hline
$m_r$ & 30 & AIC & 1 & 62 & 33 & 4 & \bf 0 & 0 & 0\\
&&BIC & 4 & 86 & 10 & 0 & \bf 0 & 0 & 0\\
&&${\rm BIC_I}$ & 0 & 28 & 31 & 24 & \bf 13 & 1 & 3\\
&50 & AIC & 0 & 27 & 65 & 7 & \bf 1 & 0 & 0\\
&&BIC & 5 & 80 & 15 & 0 & \bf 0 & 0 & 0\\
&&${\rm BIC_I}$ & 0 & 2 & 21 & 15 & \bf 42 & 12 & 8\\
&100 & AIC & 0 & 9 & 51 & 24 & \bf 16 & 0 & 0\\
&&BIC & 0 & 58 & 41 & 1 & \bf 0 & 0 & 0\\
&&${\rm BIC_I}$ & 0 & 0 & 0 & 4 & \bf 34 & 17 & 45\\
\hline
\end{tabular}
\end{center}
\end{table}
Tables \ref{tab:mm-sim1}, \ref{tab:mm-sim2} and Figure \ref{fig:num-mixedmodel} contain the results.  From these results, we find that the ${\rm BIC_I}$ selects the model which minimizes the MSE and with the correct number of basis functions.  
\section{Concluding remarks}
In this thesis we estimated the nonlinear mixed effects model by the maximum likelihood method, however, it is considered estimating it by the maximum penalized likelihood method.  The penalized log-likelihood function for the nonlinear mixed effects model (\ref{mm-pdf}) may be given by 
\begin{align*}
l_\zeta(\bm\theta)=l(\bm\theta)-\frac{n\zeta}{2}\bm\beta^T\Omega\bm\beta,
\end{align*}
where $\Omega$ is a positive semi-definite matrix and $\zeta$ is a smoothing parameter.  For the choice of $\zeta$, model selection criteria \cite{KoKi2008} will be needed, whose derivations remain as a future work.  

\cite{JaHaSu2000} extended the mixed effects model and proposed a reduced rank mixed effects model for sparse longitudinal data.  They also considered estimating it by the maximum penalized maximum likelihood method.  It is considered to derive model selection criteria for evaluating the estimated model.  
\section*{Appendix: The EM algorithm}
Steps of the EM algorithm for estimating the nonlinear mixed effects model are as follows:
\begin{description}
\item[Step 0.] Let 
$\bm\theta_{(0)} = \{\bm\beta_{(0)}^T, ({\rm vech}\Gamma_{(0)})^T, \sigma_{\varepsilon,(0)}^2\}^T$ 
be an initial value of the parameter 
$\bm\theta = \{\bm\beta^T, ({\rm vech}\Gamma)^T, \sigma_{\varepsilon}^2\}^T$.  
\item[Step 1.] (E-step) For the $j$-th iteration, calculate the conditional expectation $\bm\gamma_{\alpha,(j)}$ as follows:
\begin{align*}
\bm\gamma_{\alpha,(j)} = (\sigma^{2}_{\varepsilon,(j)} \Gamma^{-1}_{(j)} + \Phi^{r^{T}}_\alpha \Phi^r_\alpha)^{-1}\Phi^{r^{T}}_\alpha (\bm x_\alpha - \Phi^f_\alpha \bm\beta_{(j)}).
\end{align*}
\item[Step 2.] (M-step) Update $\sigma^{2}_{\varepsilon}$ as follows:
\begin{align*}
\sigma^{2}_{\varepsilon,(j+1)} =& \frac{1}{\sum N_\alpha}\sum_{\alpha=1}^n
\left[\|\bm x_\alpha - \Phi^f_\alpha\bm\beta_{(j)} - \Phi^r_\alpha \bm\gamma_{\alpha,(j)}\|^2
 + {\rm tr} \left\{\Phi^r_\alpha\left(\Gamma^{-1}_{(j)} +\frac{1}{\sigma^{2}_{\varepsilon,(j)}}\Phi^{r^{T}}_\alpha\Phi^r_\alpha\right)^{-1}\Phi^{r^{T}}_\alpha\right\}\right] .
\end{align*}
\item[Step 3.] (M-step) Update $\Gamma$ as follows:
\begin{align*}
\Gamma_{(j+1)} =\frac{1}{n}\sum_{\alpha=1}^n \left\{\bm\gamma_{\alpha,(j)}\bm\gamma_{\alpha,(j)}^{T} + \left(\Gamma_{(j)}^{-1} + \frac{1}{\sigma^{2}_{\varepsilon,(j+1)}}\Phi^{r^{T}}_{\alpha}\Phi^r_\alpha\right)^{-1}\right\}.
\end{align*}
\item[Step 4.] (M-step) Update $\bm\beta$ as follows:
\begin{align*}
{\bm\beta}_{(j+1)} = \left(\sum_{\alpha=1}^n \Phi_{\alpha}^{f^{T}}\Phi_{\alpha}^f\right)^{-1}\Phi^{f^{T}}_{\alpha}(\bm x_\alpha-\Phi_\alpha^r\bm\gamma_{\alpha,(j)}).
\end{align*}
\item[Step 5.] Continue from {\bf Step 1} to {\bf Step 4} until a suitable convergence criterion is satisfied.
\end{description}
\section*{Appendix B. Details of the matrix $I(\bm\theta)$}
We show the elements of the matrix $I(\bm\theta)$ included in ${\rm BIC_I}$ (\ref{BIC2}).
It is given as follows:
\begin{align*}
&I(\bm\theta) = -\frac{1}{n}\sum_{\alpha=1}^n \frac{\partial^2 \log f(\bm x_\alpha|\bm\theta)}{\partial\bm\theta\partial\bm\theta^{T}},
&\frac{\partial^2 \log f(\bm x_\alpha|\bm\theta)}{\partial\bm\theta\partial\bm\theta^{T}}
 = \left(\begin{array}{ccc}
I_{11}^{(\alpha)}(\bm\theta) & 
I_{12}^{(\alpha)}(\bm\theta) & 
I_{13}^{(\alpha)}(\bm\theta)\\
I_{12}^{(\alpha)T}(\bm\theta) & 
I_{22}^{(\alpha)}(\bm\theta) & 
I_{23}^{(\alpha)}(\bm\theta)\\
I_{13}^{(\alpha)T}(\bm\theta) & 
I_{23}^{(\alpha)T}(\bm\theta) & 
I_{33}^{(\alpha)}(\bm\theta)\\
\end{array}\right),\notag
\end{align*}
where 
\begin{align*}
I_{11}^{(\alpha)}(\bm\theta)=&\frac{\partial^2 \log f(\bm x_\alpha|\bm\theta)}{\partial\bm\beta\partial\bm\beta^{T}}= -\Phi^{f^{T}}_\alpha W^{-1}_\alpha \Phi^f_\alpha,\\ 
I_{12}^{(\alpha)}(\bm\theta)=&\frac{\partial^2 \log f(\bm x_\alpha|\bm\theta)}{\partial\bm\beta\partial({\rm vech}\Gamma)^{T}}~~{\rm with}\\
&\displaystyle{\frac{\partial^2 \log f(\bm x_\alpha|\bm\theta)}{\partial\bm\beta\partial\Gamma_{hk}}= -\Phi^{f^{T}}_\alpha W^{-1}_\alpha\Phi^r_\alpha(\Delta_{hk} +\Delta_{kh})\Phi^{r^{T}}_\alpha W^{-1}_\alpha \bm a_\alpha~~(h \neq k)},\\
& \displaystyle{\frac{\partial^2 \log f(\bm x_\alpha|\bm\theta)}{\partial\bm\beta\partial\Gamma_{hh}}= -\Phi^{f^{T}}_\alpha W^{-1}_\alpha\Phi^r_\alpha\Delta_{hh}\Phi^{r^{T}}_\alpha W^{-1}_\alpha \bm a_\alpha},\\
I_{13}^{(\alpha)}(\bm\theta) =& \frac{\partial^2 \log f(\bm x_\alpha|\bm\theta)}{\partial\bm\beta^T\partial\sigma^2_\varepsilon}=
 -\Phi_\alpha^{f^T} W_\alpha^{-2}\bm a_\alpha,\\
I_{22}^{(\alpha)}(\bm\theta)=& \frac{\partial^2 \log f(\bm x_\alpha|\bm\theta)}{\partial({\rm vech}\Gamma)\partial({\rm vech}\Gamma)^{T}}
~~{\rm with}\\
& \begin{array}{ll}
\displaystyle{\frac{\partial^2 \log f(\bm x_\alpha|\bm\theta)}{\partial\Gamma_{hk}\partial\Gamma}=} &
\displaystyle{\Phi_\alpha^{r^{T}}W_\alpha^{-1}\{\Phi_\alpha^r(\Delta_{hk} + \Delta_{kh})\Phi_\alpha^{r^{T}} 
- \Phi^r_\alpha(\Delta_{hk} +\Delta_{kh})\Phi_\alpha^{r^{T}}W_\alpha^{-1}\bm a_\alpha \bm a^{T}_\alpha}\\
& \displaystyle{- \bm a_\alpha \bm a^{T}_\alpha W^{-1}_\alpha \Phi^r_\alpha(\Delta_{hk} +\Delta_{kh})\Phi^{r^{T}}_\alpha\}W_\alpha^{-1}\Phi_\alpha^r}\\
& \displaystyle{- \frac{1}{2}{\rm diag}[\Phi_\alpha^{r^{T}}W_\alpha^{-1}\{\Phi_\alpha^r(\Delta_{hk} + \Delta_{kh})\Phi_\alpha^{r^{T}}} \\
&\displaystyle{- \Phi^r_\alpha(\Delta_{hk} +\Delta_{kh})\Phi_\alpha^{r^{T}}W_\alpha^{-1}\bm a_\alpha \bm a^{T}_\alpha}\\
&\displaystyle{ - \bm a_\alpha \bm a^{T}_\alpha W^{-1}_\alpha \Phi^r_\alpha(\Delta_{hk} +\Delta_{kh})\Phi^{r^{T}}_\alpha\}W_\alpha^{-1}\Phi_\alpha^r]~~(h\neq k),}\end{array}\\
& \begin{array}{ll}
\displaystyle{\frac{\partial^2 \log f(\bm x_\alpha|\bm\theta)}{\partial\Gamma_{hh}\partial\Gamma}=} &
\displaystyle{\Phi_\alpha^{r^{T}}W_\alpha^{-1}\{\Phi_\alpha^r\Delta_{hh}\Phi_\alpha^{r^{T}} - \Phi^r_\alpha\Delta_{hh}\Phi_\alpha^{r^{T}}W_\alpha^{-1}\bm a_\alpha \bm a^{T}_\alpha }\\
& \displaystyle{-\bm a_\alpha \bm a^{T}_\alpha W^{-1}_\alpha \Phi^r_\alpha\Delta_{hh}\Phi^{r^{T}}_\alpha\}W_\alpha^{-1}\Phi_\alpha^r
- \frac{1}{2}{\rm diag}[\Phi_\alpha^{r^{T}}W_\alpha^{-1}\{\Phi_\alpha^r\Delta_{hh}\Phi_\alpha^{r^{T}}}\\
& \displaystyle{- \Phi^r_\alpha\Delta_{hh}\Phi_\alpha^{r^{T}}W_\alpha^{-1}\bm a_\alpha \bm a^{T}_\alpha - \bm a_\alpha \bm a^{T}_\alpha W^{-1}_\alpha \Phi^r_\alpha\Delta_{hh}\Phi^{r^{T}}_\alpha\}W_\alpha^{-1}\Phi_\alpha^r]},
\end{array}\\
I_{23}^{(\alpha)}(\bm\theta)=&\frac{\partial^2 \log f(\bm x_\alpha|\bm\theta)}{\partial({\rm vech}\Gamma)\partial\sigma^2_\varepsilon} 
~~{\rm with}\\
& \begin{array}{ll}
\displaystyle{\frac{\partial^2 \log f(\bm x_\alpha|\bm\theta)}{\partial\Gamma\partial\sigma^2_\varepsilon}=} & 
\displaystyle{\Phi_\alpha^{r^{T}}W_\alpha^{-1}(I_{N_\alpha} -W_\alpha^{-1}\bm a_\alpha\bm a^{T}_\alpha-\bm a_\alpha\bm a^{T}_\alpha W_\alpha^{-1})W_\alpha^{-1}\Phi_\alpha^r}\\
& \displaystyle{-\frac{1}{2}{\rm diag}\{\Phi_\alpha^{r^{T}}W_\alpha^{-1}(I_{N_\alpha} -W_\alpha^{-1}\bm a_\alpha\bm a^{T}_\alpha-\bm a_\alpha\bm a^{T}_\alpha W_\alpha^{-1})W_\alpha^{-1}\Phi_\alpha^r\}},
\end{array}\\
I_{33}^{(\alpha)}(\bm\theta)=&\frac{\partial^2 \log f(\bm x_\alpha|\bm\theta)}{\partial\sigma^2_\varepsilon\partial\sigma^2_\varepsilon}=
\frac{1}{2}{\rm tr}(W^{-2}_\alpha) - \bm a^{T}_\alpha W^{-3}_\alpha\bm a_\alpha.\\
\end{align*}
Here we abbreviated $f(\bm x_\alpha|\bm t_\alpha;\bm\theta)$ to $f(\bm x_\alpha|\bm\theta)$ and used the following notations:  
\begin{align*}
W_\alpha =& \sigma^2_\varepsilon I_{N_\alpha} + \Phi^r_\alpha\Gamma \Phi^{r^{T}}_\alpha,~~\bm a_\alpha = \bm x_\alpha -\Phi^f_\alpha\bm\beta,\\
\Delta_{hk} =& \left(\Delta_{hk(i,j)}\right)_{1\le i, j\le m_r},~~\Delta_{hk(i,j)} = 
\left\{\begin{array}{cc}
1 & ({\rm if}~~ i = h, j = k)\\
0 & ({\rm otherwise}).
\end{array}\right. 
\end{align*}

\end{document}